# When Patients Go to "Dr. Google" Before They Go to the Emergency Department

Michael A. Grasso, MD, PhD[1], Alexandra Rogalski, MD[1], Naveed Farrukh, MD, MPH[2], Anantaa Kotal, PhD[3], Enrique Calleros, BS Candidate[3]

[1]University of Maryland School of Medicine, Baltimore, Maryland;
[2]The Ohio State University College of Medicine, Columbus, Ohio;
[3]University of Texas, El Paso, Texas

**Abstract**

*Approximately one-third of adults search the internet for health information before visiting an emergency department (ED), with 75% encountering inaccurate content. This study examines how such searches influence patient care. We conducted an observational study of ED visits over a 12-month period, surveying 214 of 576 patients about pre-ED internet use. Data on demographics, comorbidities, acuity, orders, prescriptions, and dispositions were extracted. Patients who searched were typically younger, healthier, and more educated. Most used a general search engine to ask symptom-related questions. Compared to non-searchers, they were less likely to receive lab tests (RR 0.78, p=0.053), imaging (RR 0.75, p=0.094), medications (RR 0.67, p=0.038), or admission (RR 0.68, p=0.175). They were more likely to leave against medical advice (RR 1.67, p=0.067) and receive opioids (RR 1.56, p=0.151). Findings suggest inaccurate health information may contribute to mismatched expectations and altered care delivery.*

**Introduction**

Previous studies have shown that approximately 80% of individuals have used the internet to seek health-related information at some point. Furthermore, roughly 35% of people have searched the internet for information on acute medical issues just prior to seeking emergency medical care.[1] Among those who conducted these searches, 50% sought information on symptoms, while one third sought information on specific diagnoses, with younger and more educated individuals exhibiting higher electronic health literacy.[2,3] Additional search topics included treatment options, prognosis, information on specific hospitals, and service availability.[4]

The increasing use of online consumer health websites raises significant concerns. The Medical Library Association has identified several reputable consumer health websites, including cancer.gov, cdc.gov, and medlineplus.gov.[5] However, despite the availability of these high-quality resources, 8 out of 10 individuals begin their health-related queries using search engines like Google, selecting websites based more on convenience than accuracy.[6] This approach often leads to unstructured query results of variable accuracy, with up to 75% of the health information from these searches containing inaccurate information, and with doctors disagreeing with internet-based diagnoses as much as 70% of the time.[1,4,7,8,9] In addition to accuracy issues, search results from consumer health websites may also be limited by relevance, timeliness, and bias. Furthermore, search engine results can be influenced by factors such as popularity, financial sponsorship, political pressure, public sentiment, and various intrinsic factors that can be readily manipulated.[10,11,12]

The inaccuracy of online consumer health information is a notable concern. This is all the more important in the setting of the COVID-19 pandemic, where online health information searches increased dramatically, and where quality gaps have been well-documented.[13,14,15,16] In light of the increasing number of individuals who turn to the internet for health information, particularly during times of national or global crises, it is crucial to ensure the accuracy of such sources and to comprehend their effect on healthcare delivery.

Previous studies have shown that numerous consumer health websites contain inaccurate information and that physicians often disagree with patient expectations derived from these sources.[1,5,6,7,8,9] To date, we have found no studies have explored the potential impact of these mismatched expectations on the delivery of healthcare services. The primary aim of our study was to investigate whether patient-initiated searches for health information online could influence the care received by individuals seeking acute medical care in the emergency department (ED). Specifically, we sought to determine whether patient-initiated searching the internet about an acute medical issue could prompt different clinical decisions and whether the information obtained from such searches - colloquially known as "Dr. Google" - could affect patient outcomes.

**Methods**

The University of Maryland Medical Center (UMMC) is a teaching hospital located in downtown Baltimore. With 757 beds, it offers comprehensive healthcare services to individuals across Maryland and the Mid-Atlantic region, with a team of approximately 1,000 attending physicians and 6,500 employees. The hospital receives over 35,000 inpatient admissions, 165,000 outpatient visits, and 53,000 ED visits annually. Additionally, the Department of Emergency Medicine operates within the University of Maryland Medical System, which includes 17 emergency departments and over 275,000 patient encounters each year.

In this observational study, we surveyed and analyzed clinical data from patients seeking acute medical care in the UMMC ED. This study was supported by resources and facilities from the University of Maryland Department of Emergency Medicine, and took place in the UMMC ED. It was supported through a grant from the University of Maryland Emergency Medical Associates. Regulatory approval for this work was obtained from the University of Maryland School of Medicine Institutional Review Board and the University of Maryland Medical Center Emergency Department Research Operations Committee.

During the 12-month study period, we recruited patients in the UMMC ED using a convenience sampling approach. Informed consent was obtained from each study participant. The inclusion criteria for the study were adult patients between 18 and 89 years of age, with a distribution of women and minorities corresponding to the demographics of the UMMC ED patient population. All data collected were stored securely within the University of Maryland School of Medicine, with access limited to study personnel who had completed the required training.

We conducted a survey of patients seeking acute medical care in the UMMC ED to identify whether they searched the internet and what types of questions they initiated. See Figure 1. The survey was administered to participants in the ED waiting room or at the bedside, after they were triaged but before their ED care was completed. The survey was conducted by several medical students, who were compensated for their work on this project. Each survey took approximately 15 minutes to complete.

---

1) What is your highest level of education?
   (Elementary School, High School, College, Graduate School)

2) What is your profession?

3) Did you search an online resource before coming to the Emergency Department?
   (Yes/No)

If the patient answered "Yes", then the following questions will also be asked.

4) Did you use your phone, table, or computer to do this search?
   (Phone, Tablet, Computer)

5) What online resource(s) did you search?
   (eg, Google, MedlinePlus. If more than one, identify primary resource.)

**Figure 1**. Patient survey questions.

---

For those patients who completed the survey, we extracted ED encounter data from the UMMC clinical repository. The data included information on demographics, comorbidities, utilization of services during the ED encounter, and patient disposition.

Demographic measures collected in the study were age, gender, race/ethnicity, and education level. The medical comorbidities tracked were coronary artery disease, diabetes mellitus, hypertension, malignancy, osteoarthritis, depression, and substance misuse, with a comorbidity score calculated as the sum of these seven conditions. ED service utilization data included the number of lab tests, number of imaging studies, medications prescribed, ED acuity, and return ED visits within 7 days. The opioids prescribed were codeine, hydrocodone, hydromorphone, fentanyl, methadone, morphine, oxycodone, and tramadol. Acute ED complaints included syncope, chest pain, arrhythmia, dyspnea, abdominal pain, acute bleeding, trauma, poisoning, and volume overload. Disposition included patients who were discharged, admitted, and those who left before the completion of the ED encounter.

We summarized the characteristics of the participants by age, gender, education, race/ethnicity, comorbidities, utilization of ED services, patient acuity, diagnosis, and disposition. The chi-squared test with 95% confidence intervals was used to examine differences among characteristics between groups and to evaluate the impact of potential confounders. The chi-squared test was selected because the data involved categorical variables, which were well-suited for this statistical method.

**Results**

During the study period, we approached 576 patients seeking acute medical care in the UMMC ED, of whom 214 (37%) agreed to participate in our survey. Our participants represented 0.4% of all ED visits during the study period, which totaled 52,852. The median age of our participants was 42 years (interquartile range 30-58), 39% were male, 29% were Caucasian, and 36% had some college education. We found that those who searched the internet before coming to the ED were younger, more educated, and had fewer comorbidities, with only age showing a significant p-value of 0.023.

The median number of comorbidities in both groups was 1 out of a possible 7, with an interquartile range of 0-2. Among the participants, 6% had coronary artery disease, 20% had diabetes mellitus, 42% had hypertension, and 21% had depression. On average, those who searched the internet before coming to the ED had fewer comorbidities than those who did not (0.92 vs. 1.28, p-value 0.048). Moreover, those who searched had a lower proportion of acute ED complaints (25% vs. 32%, p-value 0.282). These confounders were accounted for in the analysis. Demographic information for the study population can be found in Table 1.

**Table 1.** Demographic and clinical characteristics of the patient cohort.

| Characteristic | Searched | Did Not Search | P Value |
|---|---|---|---|
| Demographics | | | |
|   Number of Subjects, n (%) | 62 (29%) | 152 (71%) | |
|   Age, median years (IQR) | 37 (29-51) | 46 (30-59) | 0.023 |
|   Male, n (%) | 24 (39%) | 60 (39%) | 0.917 |
|   Caucasian, n (%) | 18 (29%) | 46 (30%) | 0.934 |
|   College Educated, n (%) | 27 (44%) | 49 (32%) | 0.117 |
| | | | |
| Comorbidity Score, average | 0.92 | 1.28 | 0.048 |
|   Coronary Artery Disease, n (%) | 1 (2%) | 11 (7%) | |
|   Diabetes Mellitus, n (%) | 12 (19%) | 30 (20%) | |
|   Hypertension, n (%) | 22 (35%) | 67 (44%) | |
|   Osteoarthritis, n (%) | 5 (8%) | 20 (13%) | |
|   Malignancy, n (%) | 5 (8%) | 18 (12%) | |
|   Depression, n (%) | 11 (18%) | 33 (22%) | |
|   Substance Misuse, n (%) | 1 (2%) | 15 (10%) | |
| | | | |
| Acute ED Complaint | 15 (25%) | 48 (32%) | 0.282 |

As shown in Table 2, the majority of participants (58 out of 62, 94%) who accessed a consumer health website prior to visiting the ED began their search using an engine like Google. The most common commercial web site searched was WebMD.com. Other sites specifically search included ClevelandClinic.org, MayoClinic.org, MedlinePlus.gov, PubMed.gov, and the UMMC patient portal. The majority of all who searched (37 participants, 60%) submitted queries about specific symptoms.

| Table 2. Consumer health searches. | |
|---|---|
| Search Characteristic | |
| Search Method, n (%)* | |
|     Search Engine | 58 (94%) |
|     Commercial Site** | 19 (31%) |
|     Government or Hospital Website | 8 (13%) |
| Search Question, n (%)* | |
|     Symptoms | 37 (60%) |
|     Diagnosis or Treatment | 34 (55%) |
|     Hospital Resources | 8 (13%) |

\* Participants may have used more than one method, which resulted in total percentages greater than 100.
\*\* Commercial sites were for-profit entities not affiliated with a government agency or medical system, the most commonly used by study participants being WebMD.

Table 3. Emergency department encounters.

| Characteristic | Searched | Did Not Search | Relative Risk | P Value |
|---|---|---|---|---|
| Utilization of ED Services, n (%) | | | | |
|   Laboratory Testing | 32 (52%) | 100 (66%) | 0.78 | 0.053 |
|   Imaging Studies | 24 (39%) | 78 (51%) | 0.75 | 0.094 |
| ED Disposition, n (%) | | | | |
|   Discharged | 33 (53%) | 83 (55%) | 0.96 | 0.788 |
|   Admitted | 12 (19%) | 43 (28%) | 0.68 | 0.175 |
|   Left Before Completion | 17 (27%) | 24 (16%) | 1.67 | 0.067 |
|   Returned to ED within 7 Days | 5 (8%) | 18 (12%) | 0.68 | 0.418 |
| Medications Prescribed, n (%) | | | | |
|   All Medications | 19 (31%) | 70 (46%) | 0.67 | 0.038 |
|   Opioid Medications | 11 (58%) | 31 (44%) | 1.56 | 0.151 |

Patients who searched an online consumer health website before presenting to the ED had a lower likelihood of undergoing laboratory testing (52% vs. 66%, relative risk 0.78, p-value 0.053). Similarly, those who searched were less likely to receive imaging studies (39% vs. 51%, relative risk 0.75, p-value 0.094).

Most participants in the study were discharged home after their ED visit (55%), with no significant differences between groups (53% who searched, 55% who did not search, relative risk 0.96, p-value 0.788). However, there were notable differences in other patient dispositions. Patients who searched an online consumer health website before presenting to the ED were less likely to be admitted (19% vs. 28%, relative risk 0.68, p-value 0.175) and had a higher rate of leaving against medical advice or leaving without being seen by a physician (27% vs. 16%, relative risk 1.67, p-value 0.067). Patients who searched were also less likely to have a return ED visit within 7 days (8% vs. 12%, relative risk 0.68, p-value 0.418).

Patients who searched an online consumer health website prior to presenting to the ED were less likely to receive medication (31% vs. 46%, relative risk 0.67, p-value 0.038). However, those who searched and received medication were more likely to receive an opioid (58% vs. 44%, relative risk 1.56). ED encounters are summarized in Table 3.

**Discussion**

A growing body of evidence highlights the limitations of online consumer health information, particularly in terms of accuracy and concordance with physician recommendations. Previous studies have primarily focused on the types of online resources utilized, levels of patient satisfaction, and predictive models related to online health information-seeking behavior.[1,2,3,4,6,17] Our study contributes to this literature by shedding light on how a lack of credible online health information may influence patient care and clinical decision-making. Further research is warranted to better understand the role of patient-initiated online searches in shaping health-related decision-making and patient outcomes.

In our study, patients who searched the internet prior to visiting the ED were generally younger, more educated, and had fewer comorbidities, with these findings consistent with prior research.[1,2,3] Approximately 29% of patients reported searching the internet for health information before their ED visit, with 94% initiating their queries via general-purpose search engines and 60% searching specifically for information related to their symptoms.

Laboratory testing and radiological studies are common components of ED care, with about 50% and 20% of patients undergoing these tests, respectively.[18] Decisions regarding diagnostic testing are influenced by factors such as their potential impact on diagnosis and treatment, cost considerations, provider experience, diagnostic complexity, and institutional protocols.[19,20,21] In our study, we found that patients who searched online before presenting to the ED were significantly less likely to undergo laboratory or imaging studies, regardless of their acuity level. This suggests that internet-influenced patient expectations may have contributed to a disconnect between patients and physicians, potentially leading to deviations from established clinical decision-making guidelines.

Additionally, our findings indicate that patients who searched the internet were less satisfied with their ED care. They were more likely to leave against medical advice and less likely to return to the ED within seven days. These observations align with previous studies that have linked leaving the ED prematurely to lower patient satisfaction.[22,23] We hypothesize that online health searches may have fostered unrealistic or mismatched expectations, contributing to dissatisfaction and early departure.

Interestingly, patients who searched online received fewer medications overall but were more likely to receive opioid prescriptions. Opioid prescribing is influenced by a complex interplay of factors, including patient expectations, contraindications to non-opioid alternatives, patient demographics, comorbidities, clinical guidelines, physician workload, and physician experience.[24,25,26] We hypothesize that misaligned expectations stemming from internet searches may have inadvertently influenced prescribing patterns, resulting in a more permissive approach to opioid prescribing in an effort to meet patient demands or alleviate dissatisfaction.

Existing literature suggests that a substantial proportion of health information obtained online is inaccurate, with error rates estimated as high as 75%, and that physicians disagree with internet-based diagnoses up to 70% of the time.[2,3,8,9,10] We propose that this discrepancy between patient-acquired information and physician expertise may have created a communication gap, ultimately affecting both clinical decision-making and patient satisfaction. Future research should aim to further explore patient motivations, expectations, and satisfaction in relation to online health information, as well as gather physician perspectives on how to navigate these encounters. Additionally, efforts should focus on developing strategies to guide patients toward reliable online health resources and improve digital health literacy.

Our study has some notable limitations. The sample size was limited to the 214 subjects who participated in our survey. The participants were limited to a single academic ED. In addition, this was an observational study that should not be interpreted as causal. Self-reported survey data may also be subject to recall bias or social desirability bias, particularly when reporting internet use or health-seeking behaviors. Finally, the study did not assess the specific content or accuracy of the information patients accessed, making it difficult to directly link misinformation to observed outcomes or mismatched expectations. However, our findings were statistically significant and accounted for several potential confounders (age, gender, education, comorbidities). In addition, our findings were consistent with past studies on patient-initiated search patterns and the accuracy of online health information.

**Conclusion**

Our results underscore concerns about the accuracy and reliability of consumer health information accessed via general search engines. Previous studies have demonstrated that much of the health information encountered online is outdated, misleading, or incorrect, which may erode the trust and shared decision-making necessary for optimal patient care. In the context of the COVID-19 pandemic, where reliance on online resources increased substantially, these issues are even more relevant. Future efforts should focus on promoting awareness of reputable health information sources and integrating digital health literacy into public health education. Furthermore, healthcare providers should be encouraged to inquire about patients' pre-visit online searches to address misconceptions and align care plans accordingly.